\documentclass{mem}
\usepackage{natbib}
\usepackage{txfonts}
\usepackage{balance}
\usepackage{graphicx}
\usepackage[a4paper]{hyperref}
\idline{0}{0}
\begin{document}

\title{On the Sensitivity of the H$\alpha$ Scattering Polarization to Chromospheric Magnetism}

\author{ Ji\v{r}\'{\i} \v{S}t\v{e}p\'an$^{1,2,3}$ \and Javier Trujillo Bueno$^{1,2,4}$ }

\offprints{J. \v{S}t\v{e}p\'an}
 
\institute{
${}^1$Instituto de Astrof\'{\i}sica de Canarias, Via L\'actea s/n,
E-38205 La Laguna, Tenerife, Spain.\\
${}^2$Departmento de Astrof\'{\i}sica, Universidad de La Laguna, Tenerife. Spain.\\
${}^3$Astronomical Institute ASCR, v.v.i., Ond\v{r}ejov, Czech Republic.\\
${}^4$Consejo Superior de Investigaciones Cient\'\i ficas, Spain.\\
\email{stepan@iac.es, jtb@iac.es}
}

\authorrunning{\v{S}t\v{e}p\'an \& Trujillo Bueno}

\titlerunning{H$\alpha$ as Diagnostics of Chromospheric Magnetism}

\abstract{
A particularly interesting line for exploring the physical conditions of the quiet solar chromosphere is H$\alpha$, 
but its intensity profile is magnetically insensitive and the small circular polarization signatures produced by the longitudinal Zeeman effect come mainly from the underlying photosphere. Here we show that the Hanle effect in H$\alpha$ provides quantitative information on the magnetism of the quiet chromosphere. To this end, we calculate the response function of the emergent scattering polarization to perturbations in the magnetic field.

\keywords{ magnetic fields -- polarization -- radiative transfer -- scattering
-- Sun: chromosphere }
}
\maketitle{}

\section{Introduction}

The only way to obtain reliable empirical information on the magnetic fields of the solar chromosphere is through the measurement and physical interpretation of the emergent polarization in chromospheric spectral lines. A particularly important line is H$\alpha$, because high-spatial resolution intensity images taken at its very line center reveal the fine scale structuring of the quiet solar chromosphere. Unfortunately, the H$\alpha$ line is so broad that in the quiet chromosphere the Zeeman effect is of little practical interest (see below). Fortunately, there is a less familiar physical mechanism that produces linear polarization in the H$\alpha$ line: scattering polarization and its modification by the Hanle effect. Here we show that the scattering polarization in H$\alpha$ provides quantitative information on the magnetic structure of the upper quiet chromosphere.

\section{Scattering polarization in H$\alpha$\label{stepan-sec:2}}

\begin{figure}
\centering
\includegraphics[width=\columnwidth]{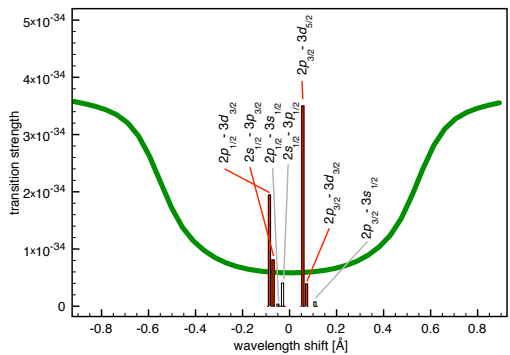}
\caption{
The H$\alpha$ line is composed of 7 fine-structure components, 4 of which (full bars) are responsible of the observed linear polarization. Compare the splitting of the components with the width of the Doppler-broadened intensity profile.
}
\label{stepan-fig:COM}
\end{figure}

H$\alpha$ in the quiet solar chromosphere is a strong absorption line which is formed under highly non-local conditions. It is a septuplet resulting from seven allowed transitions between the fine structure levels of the $n=2$ and $n=3$ levels.\footnote{Following the arguments of \citet{1982SoPh...78..157B},
the hyperfine structure is neglected in the present investigation.} The atomic terms and the fine structure levels pertaining to the same $n$ level are quasi-degenerated. Thus, the individual line components are blended due to the thermal Doppler broadening of the line (see Fig.~\ref{stepan-fig:COM}).

\begin{figure}
\centering
\includegraphics[width=\columnwidth]{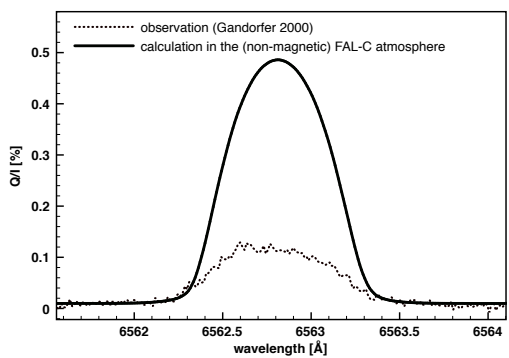}
\caption{
The observed \citep{2000sss..book.....G} $Q/I$ profile at $\mu=0.1$ vs. the calculated emergent profile in the {\em non-magnetic} atmosphere of \citet[][hereafter FAL-C model]{1993ApJ...406..319F}. Positive $Q/I$ corresponds to the tangential direction of polarization. The observed linear polarization profile represents an average profile resulting from sacrificing the spatial and temporal resolution. The disagreement between the two profiles suggests that there is a missing fundamental ingredient in such a model: the magnetic field (see text for details).
}
\label{stepan-fig:OBS}
\end{figure}

The H$\alpha$ line center is formed in the upper chromosphere while the wings originate in the photosphere \citep{1972SoPh...22..344S}.
The temperature minimum region is transparent to H$\alpha$ radiation because the population of the $n=2$ level is too small there to produce sufficient absorption in H$\alpha$. As a result, the response of the line's Stokes $V$ signal to the Zeeman effect is concentrated mainly in the photosphere of the chosen quiet atmospheric model \citep{2004ApJ...603L.129S}. Another problem with the Zeeman effect in the H$\alpha$ line is that the field strength required to produce significant Stokes-$V$ signals is too high for the quiet chromosphere. In conclusion, the Zeeman effect in H$\alpha$ is unsuitable for diagnostics of the quiet chromosphere.

Fortunately, the H${\alpha}$ line observed by \citet{2000sss..book.....G} close to the quiet solar limb shows a positive fractional linear polarization $Q/I$ profile with the maximum value at the line center (see Fig.~\ref{stepan-fig:OBS}). This scattering polarization signal results from the population imbalances and quantum coherences among the magnetic sublevels of the line's upper levels that are produced by anisotropic optical pumping processes 
\citep[eg.][]{2009ASPC..405...65T}.
While each of the 7 components of the line contributes to the absorption coefficient,  only 4 of them (those with total angular momentum of the upper level $j_u>1/2$) contribute to the emission of polarized photons. The emergent linear polarization is modified by a magnetic field via the so-called Hanle effect, which operates mainly in the line core 
\citep[eg.][]{stenflo-book1994,landi-landolfi-book-2004}. 
We point out that ``zero-field dichroism'' in the quiet chromosphere \citep{1997ApJ...482L.183T,2003PhRvL..91k1102M} is negligible for H$\alpha$ so that the 2$p_{3/2}$--3$s_{1/2}$ transition does not selectively absorb the incident radiation.

\begin{figure}
\centering
\includegraphics[width=\columnwidth]{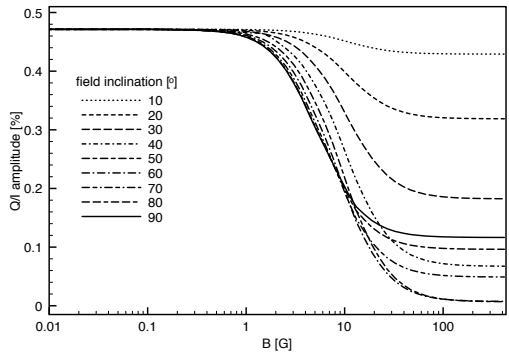}
\caption{
$Q/I$ amplitude of H$\alpha$ for a LOS with $\mu=0.1$ assuming a {\em uniformly magnetized} FAL-C atmosphere. We take into account depolarization by collisions \citep{1996A&A...309..317S}  and by a random-azimuth magnetic field with the indicated inclination.
}
\label{stepan-fig:AMP}
\end{figure}

The sensitivity of the H$\alpha$ line polarization to the magnetic field in the quiet chromosphere is investigated here by solving the multilevel non-LTE radiative transfer  problem taking into account all the relevant physical processes (anisotropic optical pumping, atomic level polarization, Hanle effect, collisional depolarization, etc.).  Fig.~\ref{stepan-fig:AMP} shows the line-center values of the calculated $Q/I$ profile for 9 different inclinations of an unresolved (i.e., with a random azimuth) magnetic field. In these model calculations, the field intensity is uniform within the whole atmosphere. Note that the line polarization is sensitive to field strengths between approximately  1 and 50 G.

\section{Response functions\label{stepan-sec:3}}

\begin{figure}
\centering
\includegraphics[width=\columnwidth]{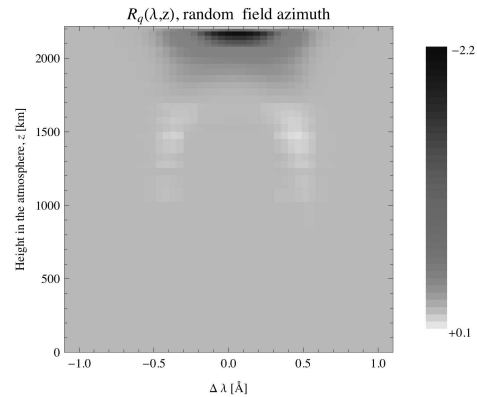}
\caption{
Response function $R_{B}(\lambda,z)$ of the emergent $Q/I$ profile of H$\alpha$ (LOS with $\mu=0.1$) to the Hanle effect of a horizontal magnetic field with a random azimuth. The unperturbed model is the non-magnetic FAL-C atmosphere.  Note that the chromosphere-corona transition region starts at 2200\,km. Dark areas indicate a decrease of the $Q/I$ signal while bright areas indicate an increase of a signal. The strong line-center response in the upper chromospheric layers makes the H$\alpha$ line particularly promising for mapping the magnetic field in the upper chromosphere of the quiet Sun.
The response function units: 10$^{-6}$\,G$^{-1}$\,km$^{-1}$.
}
\label{stepan-fig:RF}
\end{figure}

The spatio-temporally averaged observation of \citet{2000sss..book.....G} shows that at $\mu=\cos\theta\approx 0.1$ (with $\theta$ being the heliocentric angle) the amplitude of the $Q/I$ profile is about 0.12\% (see Fig.~\ref{stepan-fig:OBS}).
To obtain this $Q/I$ amplitude 
assuming uniform magnetic field in a one-dimensional model like FAL-C
we need a magnetic strength larger than $\approx 15$\,G and with an inclination higher than 40$^\circ$ (cf. Fig.~\ref{stepan-fig:AMP}). However, the assumption of a homogenous field is unrealistic and it can only provide a first rough guess about the actual magnetization of the quiet chromosphere.

In order to investigate the magnetic sensitivity of the H$\alpha$ linear polarization profile, we calculate response functions to the Hanle effect.
In particular, we investigate the sensitivity of the emergent linear polarization profile of the H$\alpha$ line to magnetic field perturbations.\footnote{Note that in the quiet chromosphere, the intensity profile of H$\alpha$ is not sensitive to the magnetic field. Thus, the response functions of $Q/I$ and of $Q$ to perturbations in the magnetic field provide similar information.} As shown below, different parts of the $Q/I$ profile are differently affected when the magnetic field is varied at a given height in the atmosphere.
The response function provides valuable information on the quantitative reaction of the emergent linear polarization profile to the local magnetic  field.

In Fig.~\ref{stepan-fig:RF}, we show the Hanle response function of the H$\alpha$ line in the (non-magnetic) FAL-C atmosphere to a perturbation in the strength of a horizontally oriented magnetic field with a random azimuth. In order to compute the response function, we have followed the same approach used by \citet{2006ASPC..354..313U} by defining the response function of $q=Q/I$ as
\begin{equation}
R_{B}(\lambda,z)=\frac 1{b}\frac{{\rm d}}{{\rm d}z}[\Delta q_z(\lambda)]\,,
\end{equation}
where $b$ is a small perturbation in the magnetic field ($b=3$\,G in our case) and $\Delta q_z(\lambda)$ is the modification of the emergent $q(\lambda)$ signal due to a perturbation applied up to the height $z$. Obviously, a key point is that to this end we apply a multilevel non-LTE radiative transfer code for the Hanle effect.

As seen in Fig.~\ref{stepan-fig:RF}, the center of the $Q/I$ profile of H$\alpha$ is strongly reduced by a magnetic field enhancement in the upper layers of the chromospherec model, while it is practically unaffected below $z\approx 1700$\,km. On the other hand, the polarization signal in the near wings at about $\pm 0.5\,\AA$ from the line center can be slightly increased by enhancing the magnetic field at about $z\approx 1500$\,km. Note that a magnetic field in the uppermost layers also affects the H$\alpha$ wings: this time it decreases the $Q/I$ signal.

A much more detailed analysis of the response function of H$\alpha$ to the Hanle effect will be presented in a forthcoming publication (\v{S}t\v{e}p\'an \& Trujillo Bueno 2009b; in preparation).

\section{Conclusions\label{stepan-sec:4}}

The Zeeman effect in H$\alpha$ is blind to the magnetic field of the upper chromosphere of the quiet Sun. Fortunately,  observations of linear polarization of the H$\alpha$ line close to the limb show a non-negligible signal in the line center. This is due to the anisotropic illumination of the hydrogen atoms in the upper chromosphere. Here we have shown that this scattering polarization signal is significantly modified by the presence of magnetic fields. The Hanle response function of the emergent $Q/I$ profile shows that the line core linear polarization observed close to the limb is exclusively sensitive to the magnetic fields in the uppermost layers of the chromosphere while the near wings can also be affected by magnetic fields in the lower chromosphere.

Careful spectropolarimetric observations and detailed radiative transfer 
modeling of the H$\alpha$ line profile can be used as a sensitive diagnostic tool of the magnetism of the quiet chromosphere, especially in its uppermost layers, providing us with valuable quantitative information on the magnetic field there (see \v{S}t\v{e}p\'an \& Trujillo Bueno 2009a; submitted).

\begin{acknowledgements}
Partial financial support by the Spanish Ministry of
Science through project \mbox{AYA2007-63881} and by the SOLAIRE network (MTRN-CT-2006-035484) is gratefully acknowledged.
\end{acknowledgements}

\bibliographystyle{aa}

\end{document}